\documentstyle[12pt]{article}

\newcommand{\EQ}{\begin{equation}}
\newcommand{\EN}{\end{equation}}

\newcommand{\ket}[1]{\left|#1\right\rangle}      

\newcommand{\bear}{\begin{eqnarray}}
\newcommand{\ear}{\end{eqnarray}}

\begin{document}

\topmargin 0pt
\oddsidemargin 5mm
\newcommand{\NP}[1]{Nucl.\ Phys.\ {\bf #1}}
\newcommand{\PL}[1]{Phys.\ Lett.\ {\bf #1}}
\newcommand{\NC}[1]{Nuovo Cimento {\bf #1}}
\newcommand{\CMP}[1]{Comm.\ Math.\ Phys.\ {\bf #1}}
\newcommand{\PR}[1]{Phys.\ Rev.\ {\bf #1}}
\newcommand{\PRL}[1]{Phys.\ Rev.\ Lett.\ {\bf #1}}
\newcommand{\MPL}[1]{Mod.\ Phys.\ Lett.\ {\bf #1}}
\newcommand{\JETP}[1]{Sov.\ Phys.\ JETP {\bf #1}}
\newcommand{\TMP}[1]{Teor.\ Mat.\ Fiz.\ {\bf #1}}
     
\renewcommand{\thefootnote}{\fnsymbol{footnote}}
     
\newpage
\setcounter{page}{0}
\begin{titlepage}     
\begin{flushright}
UFSCARF-TH-98-33
\end{flushright}
\vspace{0.5cm}
\begin{center}
{\large  Unified algebraic Bethe ansatz for two-dimensional lattice models
}\\
\vspace{1cm}
\vspace{1cm}
{\large   M.J.  Martins } \\
\vspace{1cm}
{\em Universidade Federal de S\~ao Carlos\\
Departamento de F\'isica \\
C.P. 676, 13565-905, S\~ao Carlos (S.P.), Brazil}\\
\end{center}
\vspace{1.2cm}
     
\begin{abstract}
We develop a unified formulation of the
quantum inverse scattering method  for lattice vertex models
associated to the non-exceptional 
$A^{(2)}_{2r}$, $A^{(2)}_{2r-1}$, $B^{(1)}_r$,
$C^{(1)}_r$, $D^{(1)}_{r+1}$ and $D^{(2)}_{r+1}$ Lie algebras.
We recast the Yang-Baxter algebra in terms of novel commutation
relations between  creation, 
annihilation and diagonal fields. The solution of the $D^{(2)}_{r+1}$
model is based on an interesting sixteen-vertex model which is solvable
without recourse to a Bethe ansatz.
\end{abstract}
\vspace{.2cm}
\centerline{PACS numbers: 05.50+q, 64.60.Cn, 75.10.Hk, 75.10.Jm}
\vspace{.2cm}
\centerline{December 1998}
\end{titlepage}

\renewcommand{\thefootnote}{\arabic{footnote}}

One of the main branches of theoretical and
mathematical physics is the theory of exactly solvable models.
The most successful approach to construct integrable two-dimensional
lattice models of statistical mechanics is by solving the
celebrated Yang-Baxter equation \cite{YB}. Given a solution of this
equation, depending on a continuous parameter $\lambda$, one can define
the local Boltzmann weights of a commuting family of transfer matrices
$T(\lambda)$. A complete understanding of these models should of course
include the exact 
diagonalization of the transfer matrices, which can
provide us
non-perturbative information about the on-shell physical properties
such as free-energy thermodynamics and quasi-particle excitation behaviour.

The structure of the solutions of the Yang-Baxter equation based on
simple Lie algebras is by now fairly well understood \cite{BD}. In particular,
explicit expressions for the $R$-matrices related to non-exceptional
affine Lie algebras were exhibited in ref. \cite{JB}. Since then, many
other $R$-matrices associated 
to higher dimensional representations of these algebras have also been
determined \cite{GO}. A
long-standing  open problem in this field, except for the $A_r^{(1)}$ 
algebra \cite{MG}, is the diagonalization of their transfer matrices 
by a first principle approach, i.e. through the quantum inverse scattering
method \cite{QIS,QIS1}. This technique gives us information on the
nature of the eigenvectors, which is crucial in the investigation of the
off-shell properties such as correlators of physically relevant
operators \cite{KO}. This fact becomes even more clear thanks to 
new recent developments in the calculation of form-factors for
integrable models in a finite lattice \cite{BA,MAL}.

In this work, we offer the basic tools to solve the remaining vertex
models based on the non-exceptional Lie algebras within the quantum
inverse scattering framework. Specifically, we present a universal 
formula for the eigenvectors in terms of the creation fields and
the fundamental $R$-matrix elements of the 
$A^{(2)}_{2r}$, $A^{(2)}_{2r+1}$, $B^{(1)}_r$,
$C^{(1)}_r$, $D^{(1)}_{r+1}$ and $D^{(2)}_{r+1}$ models. This general
construction extends previous work by the author and Ramos \cite{MR}
and it is crucial in order to accommodate the solution of the twisted
$D_{r+1}^{(2)}$ model. It turns out that this solution still 
depends on the diagonalization of a sixteen-vertex model having 
fine-tuned Boltzmann weights. Interesting enough, this latter 
problem is resolved 
without recourse to a lattice Bethe ansatz.

One basic object in the quantum inverse
scattering method is the monodromy operator 
${\cal T}(\lambda)$
whose trace
over an auxiliary space ${\cal A} $ gives us the transfer matrix, 
$T(\lambda)= Tr_{{\cal A}}[{\cal T}(\lambda)]$. A sufficient 
condition for integrability
is the existence of an invertible matrix $R(\lambda,\mu)$ satisfying
the following relation
\EQ
R(\lambda,\mu) {\cal T}(\lambda)  {\otimes} {\cal T}(\mu) =
{\cal T}(\mu)  {\otimes} {\cal T}(\lambda)  R(\lambda,\mu)
\EN
where the matrix elements 
$R^{\beta_1,\beta_2}_{\alpha_1,\alpha_2}(\lambda,\mu)$ of the $R$-matrix
defined on the tensor space $ {\cal A} \otimes {\cal A} $ are c-numbers.
For the models we are going to discuss in this paper, the $R$-matrix
depends only on the difference of the rapidities $\lambda$ and $\mu$. 

As a first step in this program, one may try to construct from
the intertwining relation (1) convenient commutation rules for the
matrix elements of the monodromy matrix, which in turn can 
inspire us about the 
physical content of such elements. There is no known recipe to perform
this task, but it certainly begins with an appropriate representation
for $ {\cal T}(\lambda)$ itself. An important input is the reference
state $\ket{0}$ one uses to build up the eigenvectors 
of the transfer matrix $T(\lambda)$. If we choose $\ket{0}$ as the
highest weight state for these algebras we soon realize, 
from the properties of ${ \cal T}(\lambda) \ket{0}$, that a 
promising ansatz for the monodromy should be \cite{MR}
\EQ
{\cal T}(\lambda) =
\pmatrix{
B(\lambda)       &   \vec{B}(\lambda)   &   F(\lambda)   \cr
\vec{C^{*}}(\lambda)  &  \hat{A}(\lambda)   &  \vec{B^{*}}(\lambda)   \cr
C(\lambda)  & \vec{C}(\lambda)  &  D(\lambda)  \cr}
\EN

Here the vector
$ \vec{B}(\lambda)$ and the scalar field $F(\lambda)$ will play the role
of creation operators with respect to the reference state $\ket{0}$.
The field 
$ \vec{B}(\lambda)$ is a $(q-2)$-component $row$ vector, where $q$ is the
number of states per bond of these vertex models in a square lattice.
Its relation to the rank of each non-exceptional Lie algebra discussed
in this paper is given in table 1. 
The operator $\vec{B^{*}}(\lambda)$ represents $(q-2)$-
component $column$ vector operator, playing the role as a
redundant creation field and therefore does not enter 
in our construction of the eigenvectors. The scalar field $C(\lambda)$,
the column and row vectors 
$\vec{C^{*}}(\lambda)$ and
$\vec{C}(\lambda)$ are annihilation operators,  $\hat{A}(\lambda)$ is
a $(q-2) \times (q-2)$ matrix operator while the remaining fields
$B(\lambda)$ and $D(\lambda)$ are diagonal scalar operators. Putting 
them together, we have a rather specific $q \times q$ matrix
representation for the
monodromy matrix.

Taking into account this discussion and following the steps of
ref. \cite{MR} one can find the appropriate set of fundamental
commutation rules between the creation, annihilation and 
diagonal fields. However, in order to accommodate the solution of
the $D_{r+1}^{(2)}$ vertex model, we lead to generalized expressions
for the commutation rules as compared to those exhibited in ref. \cite{MR}.
For sake of simplicity we illustrate these modifications only in the
simplest case. This turns out to be the commutation rule between the fields
$B(\lambda)$ and
$\vec{B}(\lambda)$, which is given by 
\EQ
B(\lambda)\vec{B}(\mu) = 
 w_1(\mu-\lambda) \vec{B}(\mu)B(\lambda) - 
\hat{\eta}(\mu-\lambda). \vec{B}(\lambda)B(\mu)
\EN

For the $D_{r+1}^{2}$ model $\hat{\eta}(\lambda,\mu)$ is the
following  matrix
Boltzmann weight 
\EQ
\hat{\eta}(x) =
\pmatrix{
\hat{I}w_2(x)       &   0   &   0 & 0   \cr
0  &  w_3^{-}(x)   &  w_3^{+}(x) & 0 \cr
0  & w_3^{+}(x)  & w_3^{-}(x) & 0  \cr
0  &  0  & 0  & \hat{I}w_2(x)  \cr} 
\EN
while for the other non-exceptional Lie algebras is just a scalar 
$\hat{\eta}(x)=w_2(x)$. The operator $\hat{I}$ denotes the 
$(q/2-2) \times (q/2-2)$ Identity matrix and the expressions for the
weights $w_i(x)$ are
\EQ
w_1(x)= \frac{\exp(\alpha x) -k^2}{k[\exp(\alpha x)-1]},~~
w_2(x)= \frac{1 -k^2}{k[\exp(\alpha x)-1]},~~
w_3^{\pm}(x)= \frac{k^2-1}{2k[1 \pm \exp(\alpha x/2)]}
\EN
where $\alpha=2$ for the $D_{r+1}^{(2)}$ model and $\alpha=1$ for the other
non-exceptional Lie algebras listed in table 1. The parameter $k$ describes
the ``quantum'' deformation as defined by Jimbo \cite{JB}. We remark that
many other commutation rules need similar modifications, but since they are
sufficiently cumbersome we shall present these technical
details elsewhere \cite{MA}.

We next turn to the analysis of the structure of the eigenvectors. These 
are multiparticle states  characterized by a set of rapidities which
parametrize the creation fields, and can be written as linear
combination of products of the operators 
$\vec{B}(\lambda)$ and $F(\lambda)$ acting on the reference state $\ket{0}$.
The following physical picture helps us to construct an educated
ansatz for such multiparticle states. The field 
$\vec{B}(\lambda)$ plays the role of a single particle excitation 
while $F(\lambda)$ describes a pair excitation,
both with bare momenta parametrized by
$\lambda$. Furthermore, the total number of particles is a conserved
quantity thanks to an underlying $U(1)$ invariance. Consequently, 
$\vec{B}(\lambda_1)$ will represent the one-particle state, the linear
combination 
$\vec{B}(\lambda_1) \otimes \vec{B}(\lambda_2) 
+\vec{v}(\lambda_1,\lambda_2) F(\lambda_1)$ will be the two-particle state
for some unknown vector
$\vec{v}(\lambda_1,\lambda_2)$ and so forth. Adapting the steps of
ref. \cite{MR} to include the $D_{r+1}^{(2)}$ structure, we find after
a tedious computation that the $n$-particle eigenvector 
$\ket{\Phi_{n}(\lambda_{1}, \cdots ,\lambda_{n})}$ can be written by the
linear combination
\EQ
\ket{\Phi_{n}(\lambda_{1}, \cdots ,\lambda_{n})} =  
\vec {\Phi}_{n}(\lambda_{1}, \cdots ,\lambda_{n}).\vec{\cal{F}} \ket{0}
\EN
where the $(q-2)^{n}$-components of the vector $\vec{\cal {F}}$ describe
the linear combination and the vector
$\vec {\Phi}_{n}(\lambda_{1}, \cdots ,\lambda_{n})$ satisfies the 
following recurrence relation
\bear
\vec {\Phi}_{n}(\lambda_{1},\cdots,\lambda_{n}) = 
\vec {B}(\lambda_{1}) \otimes \vec {\Phi}_{n-1}(\lambda_{2}, 
\cdots,\lambda_{n})
- 
\sum_{j=2}^n 
\vec{\xi}(\lambda_{1}-\lambda_{j})
\prod_{k=2,k \neq j}^{n} 
w_1(\lambda_{k}-\lambda_{j}) F(\lambda_1) \otimes
\nonumber \\
\times 
 \vec {\Phi}_{n-2}(\lambda_{2},\cdots,\lambda_{j-1},\lambda_{j+1},\cdots,\lambda_{n}) B(\lambda_j) \prod_{k=2}^{j-1}
\hat{r}_{k,k+1}(\lambda_{k}-\lambda_{j})
\ear

In this formula, the vector
$\vec{\xi}(x)$ plays the role of a generalized exclusion principle,
projecting out certain forbidden states made by the creation fields
$\vec {B}(\lambda_{i})$ from the linear combination. This exclusion rule
is governed by the non-null components of this vector,
which have been determined in terms of the original $R$-matrix
elements by
\EQ
\vec{\xi}(x)= \sum_{i,j=1}^{q-2}\frac{R_{1q}^{i+1,j+1}(x)}{R_{1q}^{q1}(x)}
\hat{e}_i \otimes \hat{e}_j
\EN
where $\hat{e}_i$ denots the elementary projection on the $i$th position.

The meaning of the auxiliary $R$-matrix $\hat{r}(x)$ in the
expression (7) is that it dictates the symmetry of the eigenvectors 
under permutation of rapidities, namely
\EQ
\vec {\Phi}_{n}(\lambda_{1},\cdots,\lambda_j,\lambda_{j+1},\cdots,\lambda_{n}) = 
\vec {\Phi}_{n}(\lambda_{1},\cdots,\lambda_{j+1},\lambda_{j},\cdots,\lambda_{n})
.\hat{r}_{j,j+1}(\lambda_j-\lambda_{j+1})
\EN

As long as $r > 2$, the structure of the matrix $\hat{r}(x)$ is based on
the same Lie algebra as the original $R$-matrix we started with, but now
having a lower level rank $r-1$. This structure, however, can change
drastically when we reach the lowest level, and for this reason it
is convenient to call them separately by $\hat{r}^{(1)}(x)$ matrices.
In table 1 we describe the type of vertex models that are associated
to these $\hat{r}^{(1)}$-matrices for each non-exceptional
Lie algebra. We note that for most
of the models the underlying $\hat{r}^{(1)}$-matrices are based on
well known six \cite{LI,QIS} and nineteen vertex models \cite{FZ,IK}.
The $D_{r+1}^{(2)}$ model is however an exception and its
fundamental $\hat{r}^{(1)}$-matrix is given by a rather
peculiar sixteen-vertex
model. Since this result seems to be new in the literature, we shall
present here details about this theory, beginning with its matrix form
\EQ
\hat{r}^{(1)}_{D_2^2}(x) =
\pmatrix{
a_{+}(x,k)  &  -b(k)  & -b(k) & c(k)   \cr
b(k)  &  -c(k)  &  a_{-}(x,k) & b(k) \cr
b(k)  & a_{-}(x,k)  & -c(k) & b(k)  \cr
c(k)  &  -b(k)  & -b(k)  & a_{+}(x,k)  \cr} 
\EN
where the Boltzmann weights $a_{\pm}(x,k)$, $b(k)$ and $c(k)$ are given by
\EQ
a_{\pm}(x,k)= f_{\pm}(k) +g(x),~~ c(k)=\frac{2b^2(k)}{f_{+}(k)-f_{-}(k)}
\EN

This matrix is factorizable for arbitrary functions $f_{\pm}(k)$ and $g(x)$,
but in the specific case of the $D_{r+1}^{(2)}$ model we have the following
extra constraints
\EQ
g(x)=\sinh(x),~~ f_{+}(k)=-f_{-}(k),~~ f_{+}(k)=c(k) \pm(k-1/k)/2
\EN

To make further progress for the algebraic Bethe ansatz solution
of the $D_{r+1}^{(2)}$ it is necessary to diagonalize the auxiliary
transfer matrix associated to the $\hat{r}^{(1)}$-matrix (10) in the
presence of inhomogeneities. More precisely, we have to tackle the
following eigenvalue problem
\EQ
{\hat{r}^{(1)}}(\lambda-\mu_{1})_{b_{1}d_{1}}^{c_{1}a_{1}}
{\hat{r}^{(1)}}(\lambda-\mu_2)_{b_{2}c_{2}}^{d_{1}a_{2}} \cdots
{\hat{r}^{(1)}}(\lambda-\mu_n)_{b_{n}c_{1}}^{d_{n-1}a_{n}}
{\cal{F}}^{a_{n} \cdots a_{1}} = \Lambda^{(1)}_{D^2_2}(\lambda,\{ \mu_j \})  
{\cal{F}}^{b_{n} \cdots b_{1}}
\EN
where $\{ \mu_j \}$ stands for the inhomogeneities. 

We solve this problem by first mapping the sixteen-vertex model to an
asymmetric eight-vertex model, following a procedure devised long ago 
by Wu \cite{WU}. This leads us to a much simpler vertex model, having
the following $R$-matrix
\EQ
\hat{r}^{(1)}_{8v}(x) =
\pmatrix{
\tilde{a}(x)  &  0  & 0 & \tilde{d}_{+}(k)   \cr
0  &  \tilde{c}(k)  &  \tilde{a}(x) & 0 \cr
0  & \tilde{a}(x)  & \tilde{c}(k) & 0  \cr
\tilde{d}_{-}(k)  &  0  & 0  & \tilde{a}(x)  \cr} 
\EN
where the respective weights are given by
\EQ
\tilde{a}(x)=\sinh(x),~~ \tilde{c}(k)=f_{+}(k)-b^2(k)/f_{+}(k),~~
\tilde{d}_{\pm}(k)= f_{+}(k) +b^2(k)/f_{+}(k) \pm 2b(k)
\EN

The
Boltzmann weights of this eight-vertex model have
enough special properties to allow us exact diagonalization without the
need of a Bethe ansatz analysis. In fact, the off-diagonal matrix
component of
the corresponding Lax operator commutes due  to the relation 
$d_{+}(k)d_{-}(k)=c^2(k)$, which is also a 
restriction for factorization of the $r$-matrix (14).
This leads us to conclude that all the
eigenvectors are given in terms of
products of on site states, and  its expression 
in the spin-1/2 $\sigma^{z}$ basis is
\EQ
\vec{{\cal F}}_{8v}= \prod_{\epsilon_i=\pm 1} \prod_{i=1}^{m} \otimes 
\pmatrix{
1 \cr
\epsilon_i \sqrt{\frac{\tilde{c}(k)}{\tilde{d}_{+}(k)}} \cr}
\EN
where $\epsilon_i=\pm 1$ are $Z_2$ variables, parametrizing the 
many possible $(2)^m$
states. The expression for the eigenstates of the
original sixteen-vertex model can be obtained from (16) after
a transformation to the representation where $\sigma^{x}$ is
diagonal. It is interesting to note that these are typical
variational states, with the advantage of
being exact and valid for the role spectrum.

Now the eigenvalues can be determined almost directly, and they are
given by
\EQ
\Lambda^{(1)}_{D^2_2}(\lambda,\{ \mu_j \}) = 
\prod_{i=1}^{m} \left [ \sinh(\lambda-\mu_i) -\epsilon_i(k-1/k)/2 \right ]
+
\prod_{i=1}^{m} \left [ \sinh(\lambda-\mu_i) +\epsilon_i(k-1/k)/2 \right ]
\EN

These latter results are fundamental in order to solve the eigenvalue
problem for the $D^{(2)}_{r+1}$ from first principles. In particular,
the fact that the eigenvectors do not depend on the rapidities is an
essential feature to match the inhomogeneous eigenvalue and nested
Bethe ansatz
problems, since the $\hat{r}^{(1)}$-matrix can not be made regular. It
should be also emphasized that this theory is not in the class of
the so-called free-fermion models.

In summary, we have developed a framework which is  capable to deal
with the transfer matrix eigenvalue problem of the vertex models 
based on non-exceptional Lie algebras from a unified point of view.
Our nested Bethe ansatz results for the eigenvalues and Bethe ansatz
equations
corroborate those conjectured in ref. \cite{RE} by means of
analyticity assumptions. The many technical details of such nested Bethe
ansatz analysis will be presented elsewhere \cite{MA}. 
The universal formula we have obtained for the eigenvectors
pave the way to a general off-shell Bethe ansatz formulation, and
consequently could be useful to produce integral representations
for the form-factors \cite{BA}. Finally, we remark that all the models
solved in this work share a common algebraic structure, i.e the
braid-monoid algebra \cite{JO,WA}. A question that promptly arises
is if there are direct connections between the braid-monoid algebra
and our algebraic Bethe ansatz framework. In this sense, we note
that the size(three) of the matrix representation for ${\cal T}(\lambda)$
coincides with the number 
of eigenvalues of the braid operator \cite{JO}.  The same  observation
also works for the Hecke algebra, where the
number of eigenvalues is two \cite{JO}. It remains to be seen whether this
is an isolated coincidence or the tip of an iceberg.

\section*{Acknowledgements}
This work was partially supported by Cnpq and Fapesp.

\newpage
\underline{Table 1}: Parameters of the vertex models 
associated to the affine Lie algebras. The symbols IK and FZ
stand for  Izergin-Korepin \cite{IK} and 
Fateev-Zamolodchikov models \cite{FZ},
respectively.

\begin{table}
\begin{center}
\begin{tabular}{|c|c|c|} \hline
  Lie algebra     & $ q$ & $\hat{r}^{(1)}$-matrix   \\ \hline\hline
$A_{2r}^{(2)}$ & $2r+1$ & nineteen-vertex IK model  \\ \hline
$A_{2r-1}^{(2)}$ and $C_{r}^{(1)}$ & $2r$ & six-vertex model \\ \hline
$B_{r}^{(1)}$ & $2r+1$ & nineteen-vertex FZ model  \\ \hline
$D^{(1)}_{r+1}$ & $2r+2$ & two decoupled six-vertex models  \\ \hline
$D^{(2)}_{r+1}$  & $2r+2$ & sixteen-vertex model \\ \hline
\end{tabular}
\end{center}
\end{table}

\end{document}